\documentstyle[12pt]{article}
\newfont{\feff}{cmti10}
\def\undertext#1{\vtop{\hbox{#1}\kern 1pt \hrule}}

\def\tr{\hbox{tr}\,}

\def\be{\begin{equation}}
\def\ee{\end{equation}}
\def\bea{\begin{eqnarray}}
\def\eea{\end{eqnarray}}
\def\eqref#1{(\ref{#1})}

\begin{document}

\begin{titlepage}
\title{
\begin{flushright}
{\bf\normalsize   PUPT-1322}\\
{\bf\normalsize   LPTENS-92/15}\\
\end{flushright}
Induced QCD at Large N}

\author{
V.A. Kazakov\\
Laboratoire de Physique Theorique\\
de L'Ecole Normale Superieure, 24 rue Lhomond,\\
75231 Paris CEDEX 05, France\\
and\\
A.A. Migdal\\
Physics Department,\\
Jadwin Hall, Princeton University\\
Princeton, NJ 08544-1000\\
}
\date{\today}

\maketitle

\begin{abstract}

We propose and study at large N a new lattice gauge model , in which the
Yang-Mills interaction is induced by the heavy scalar field in adjoint
representation. At any dimension of space and any $ N $ the gauge fields can be
integrated out  yielding an effective field theory for the
 gauge invariant scalar
field, corresponding to eigenvalues of the initial matrix field. This field
develops the vacuum average, the fluctuations of which describe the elementary
excitations of our gauge theory.
At $N= \infty $ we find two phases of the model, with asymptotic freedom
corresponding to the strong coupling phase (if there are no phase
transitions at
some critical $N$). We could not solve the model in this phase, but in the weak
coupling phase we have derived  exact nonlinear integral
 equations for the vacuum
average and for the scalar excitation spectrum. Presumably the strong coupling
equations can be derived by the same method.

\end{abstract}
\end{titlepage}

\newpage

\section{Introduction}

The complexity of QCD comes from the gauge field, which is neither completely
random, nor completely classical. It is classical at small scales, according to
asymptotic freedom, and random at large ones, according to quark confinement.
When the large $ N $ approximation was invented in the seventies, we expected
something to become classical at last, but such a master field was
 never found in
spite of all efforts. At present the chance of solving large $ N $ QCD
by a WKB approximation doesn't appear promising.

Still, to our own surprise, we have found the master field.
It becomes manifest, when
we treat the gauge theory as induced quantum theory, in the spirit of Sakharov.
The details are important here, it does matter what regularization to take, and
what field to take for induction.

We choose the scalar field in adjoint representation $ \Phi(x) $ whose
eigenvalues $ \phi_a(x) $ will later serve as a master field. As for the
angular
matrices $ \Omega(x) $ diagonalizing the matrix $ \Phi(x) $, those decouple
after integrating out the gauge field.

Our theory being nonperturbative, we have to specify the regularization. We
have
not invented anything new here, just taken the standard lattice theory,
 except we
did not add the gauge field selfinteraction to our action

\begin{equation}
S =  \sum_x N \, \tr \left( m_0^2 \Phi^2(x) - \sum_{\mu= 1, 2,
..,  D}\Phi(x)U_{\mu}(x)\Phi(x+\mu) U_{\mu}^{\dagger}(x) \right)\;;\; \tr
\Phi \equiv 0\;;\
\label{Action}
\end{equation}

The parameter $ m_0 $ is not quite the bare mass, but it could play
the same role of adjusting the physical mass scale. In the next Section 2 we
 analyze
the relation of this model to the conventional QCD, and obtain a scaling
 relation
for the glueball mass scale as a function of $ m_0 $.

Then, in Section 3 we integrate out the gauge fields at every link, which
yields effective interaction for the eigenvalues $ \phi_a(x) $. This
interaction is essentially nonpolynomial, which is our only protection against
the Gaussian fixed point for the master field.
Investigating this interaction, we  find a surprising relation between the well
known Itzykson-Zuber integral over the unitary group and the two-matrix model,
which  describes the complete $ (p,q) $ table of conformal fields in 2D Quantum
Gravity. Using  modern large N technology, which is especially simple here,
as there is no need for the double scaling limit, we find exact integral
equation describing the weak coupling phase of the Itzykson-Zuber
 integral at given
density of the eigenvalues of the scalar fields.

The strong coupling phase is studied in Section 4. We  could not find
an exact analytic solution here, but rather studied the
 strong coupling expansion
which we developed up to $ m_0^{-11} $. The general method for computation to
any order of the strong coupling expansion is briefly described.

In Section 5 we derive the classical equation for the
 vacuum average of the
density of eigenvalues, and study it together with previous equations. As we
find, one of equations can be solved exactly, which leaves us with only one
nonlinear integral equation. We suggest that the critical phenomena  arise when
one of the endpoints of  the spectrum goes to zero, so that two branch cuts of
the weak coupling phase merge into one cut of the strong coupling phase.

In Section 6 we derive the effective Lagrangean for the scalar fields in
the next large $ N $ approximation, corresponding to the Gaussian distribution
for the vacuum fluctuations of eigenvalues. The corresponding quadratic
functional diagonalizes in momentum space, but remains nonlocal in eigenvalue
space. The mass spectrum is given by a certain linear integral wave equation,
depending upon the master field. As one would expect
here are in general infinitely many masses. Unfortunately though, in
the weak coupling phase we studied, the hopping term in effective wave
equation, producing the Laplace operator in continuum limit, vanishes
in the large $ N $ limit by dynamical reasons.

Of course, all these computations for the large $N$ will have some
physical meaning if there is no phase transitions happening on the way
for some finite $N$, as it happens, in the Wilson type lattice QCD
at $N=4$. But even if this unwanted phase transition would appear in our
model, there will be left still some possibilities in it to avoid it
by an appropriate choice of the (nonlinear) potential for
the field $\Phi$.

The remaining problems, and future directions are discussed in the last Section
7. We suggest numerical experiments to test its correspondence with QCD.

\section{Induced QCD}

Let us integrate out the scalar field \footnote{In this Section we include the
trace of $\Phi $ as dynamic variable, which does not affect the leading order
of the large $ N $ expansion.} in the functional integral of our lattice theory
to induce the effective gauge field action
\begin{equation}
\int DU D\Phi \exp(-S) \propto \int DU \exp\left(-S_{ind}[U]\right)
\label{Induced}
\end{equation}
The $ \frac{1}{m_0} $ expansion of the induced action
\begin{equation}
S_{ind}[U] =   \frac{1}{2} Tr \ln \left(\delta_{x,y} - m_0^{-2}\sum_{\mu}
U_{\mu}(x)\otimes U^{\dagger}_{\mu}(x) \delta_{x+\mu,y} \right)
\label{1mexpansion}
\end{equation}
can be represented as a sum over lattice
 loops $ \Gamma $ of a scalar particle in
external gauge field
\begin{equation}
S_{ind}[U] = - \frac{1}{2}\sum_{\Gamma} \frac{\left|tr U[\Gamma]\right
|^2}{l[\Gamma] m_0^{2 l[\Gamma]}}
\label{RandomLoops}
\end{equation}
where $ l[\Gamma] $ stands for the length of the loop, and $ U[\Gamma] $ for
the ordered loop product of $U$ matrices.

For weak smooth gauge fields the critical value of the bare mass would be at $
m_c^2 =  d$, but should we average over short wavelength fluctuations of the
gauge fields, this critical value might shift to some other value $m_c^2 $.
Near this value, we could take the continuum limit
 for the smooth part of the gauge
field.

Expanding  in powers of the gauge potential, we would find the usual sum of one
loop lattice Feynman graphs. It would be instructive to compute them
explicitly, without first going to the local limit in the lattice action $ S $
and switching from the lattice to dimensional regularization. Let us outline
this calculation.

We represent $ U_{\mu}(x) = \exp( a A_{\mu}(x) ) $ and choose the Schwinger
gauge
\begin{equation}
A_{\mu}(x) = -\frac{1}{2} F_{\mu \nu} x_{\nu} + \dots
\label{Schwinger}
\end{equation}
where dots stand for the higher derivatives of the field strength. Note that
the Abelian part of $ A_{\mu} $ drops from our effective action after taking
absolute value of the loop product. So, this $ F_{\mu \nu} $ is the $ SU(N) $
rather then $ U(N) $ field strength, as it should be.

The leading terms in the adjoint loop product in the local limit
\begin{equation}
\left|\tr U[\Gamma]\right|^2 \rightarrow N^2 -  \frac{N}{4} \tr \left(F_{\mu
\nu} F_{\alpha \beta}\right) \sum_{x,y \subset \Gamma} \Delta x_{\alpha} \Delta
y_{\mu} \left(x_{\beta}-y_{\beta}\right)\left(x_{\nu}-y_{\nu}\right)
\label{Fmunu2}
\end{equation}
where $ \Delta x = x^{(i)} - x^{(i-1)}$ denotes the difference between  two
consequent points on the lattice loop. The identities  $ \sum \Delta x =0 $
were utilized in this relation. Now it is a straightforward exercise in
lattice field theory to compute the sums over loops
\begin{equation}
\frac{N}{8}\tr \left(F_{\mu \nu} F_{\alpha \beta}\right)\sum_{\Gamma}
\frac{1}{l[\Gamma]m_0^{2 l[\Gamma]}}\sum_{x,y \subset \Gamma} \Delta x_{\alpha}
\Delta y_{\mu} \left(x_{\beta}-y_{\beta}\right)\left(x_{\nu}-y_{\nu}\right)
\label{LatticeLoop}
\end{equation}
The sums over points $x,y \subset \Gamma $ should be interchanged with the sums
over lattice loops $\Gamma $, which yields the sum over all $x,y $ on the
lattice times the sum over numbers of links in the two parts $
\Gamma_{xy}, \Gamma_{yx} $ of the original loop $ \Gamma $. As a result,
 the factor
$ l[\Gamma] $ in the denominator cancels, and we arrive at independent sums
over
lattice paths $ \Gamma_{xy}, \Gamma_{yx}$.

One could either go to momentum space and represent $ x-y $ as derivatives with
respect to momenta, or stay in coordinate space, and use the asymptotic form of
lattice propagator
\begin{equation}
\sum_{\Gamma_{x,y}} m_0^{-2 l\left[\Gamma_{x,y}\right] } =
m_0^2\int_{-\pi}^{\pi} \frac{d^dp}{(2 \pi)^d} \frac{\exp (\imath p
(x-y))}{m_0^2 - \sum_{\mu>0} \cos(p_{\mu})} \rightarrow d
\int_{-\infty}^{+\infty}\frac{d^dp}{(2 \pi)^d}\frac{\exp (\imath p (x-y))}{
p^2}
\label{Propagator}
\end{equation}
which is proportional to $  |x-y|^{2-d} $.

In the local limit we obtain the Yang-Mills action, with the bare coupling $
\frac{1}{g_0^2} $ given by the scalar one loop diagram with the ultraviolet
cutoff $\Lambda = a^{-1}$ , and the infrared one at the renormalized mass $ m^2
= m_0^2-m_c^2 $. In four dimensions, with proper normalization
\begin{equation}
\frac{1}{ g_0^2} \rightarrow \frac{N}{96 \pi^2}\ln \frac{\Lambda^2}{m^2}
\label{ScalarLoop}
\end{equation}

The terms with higher powers of the Yang-Mills field strength would enter with
inverse powers of $m$ , but they would not depend upon the ultraviolet cutoff,
as it follows from dimensional counting. This means that in the local limit
$ \Lambda  \rightarrow \infty $ one could safely neglect these higher terms.

This calculation does not include the feedback  of hard gluons. Now,
once we obtained the induced coupling constant for the Yang-Mills
theory, we could in principle estimate these effects. We should go
backwards, from  $ m $ to $ \Lambda $, and take into account all the
relevant interactions, including the  $ \phi^4 $ induced by gluons.
The resulting running gluon constant should blow up at $ \Lambda
$ which would yield correct estimate for the relation between $ m$ and
$ g_0^2$. This relation depends upon unspecified bare $ \phi^4 $
coupling, so that there seems to be no universality.

We could have written from the very beginning the continuous version of
our model which corresponds to the gauged scalar massive matrix field in
the adjoint representation, without an explicit Yang-Mills term:

\begin{equation}
L= \frac{N}{g_0^2} \tr  \left(\left(\partial_{\mu} \Phi +
i \left[A_{\mu},\Phi\right]\right)^2 + m^2 \Phi^2  + \lambda_0 \Phi^4\right)
\label{Contin}
\end{equation}

By the naive change of the gauge we can gauge out the angular degrees
of freedom of the $\Phi$ field here. Than the gaussian integration over
the gauge potentials leeds to an effective action for  only the
eigenvalues of $\Phi$. But this action will suffer from the obvious
drawbacks: different eigenvalues will be strongly attractive because of
the Vandermonde determinants in the negative powers in the effective
measure, which appear after the integration over vector potentials.
Another problem is that from the very beginning we miss the diagonal
degrees of freedom of the gauge field (it enters only in the
commutator), which is of course bizarre. The resolution of all these
problems will be found in the lattice version of the model where it
will be obvious that some important contributions to the functional
integral are missing in the continuous version, but are present on the
lattice.

Each of the terms in the lattice sum over loops has the same
structure, as the usual Wilson action in adjoint representation. The
fact that loops $ \Gamma $ are not elementary plaquettes, does not
seem to be important, as this was the ambiguity of the Wilson theory
anyway.

It is important though, that these loops are small, which corresponds
to heavy inducing fields. We are going to adjust the scalar mass so,
that these loops would become much larger then the lattice spacing.
In this case the effective action would be large, since the sum over
loops would be close to divergency, which would provide us with
nesessary large value of the bare lattice coupling $ \beta =
\frac{1}{g_0^2} $ of induced gauge theory.
On the other hand, the loops would be much smaller then the physical
scale, so that their size would serve as an effective cutoff
for induced
QCD.

One could go one step further in analogy with Wilson QCD. Namely, we
could construct the Wilson-like strong coupling expansion for this
multiloop gauge theory. The generic term of the Wilson strong coupling
expansion corresponds to the closed surface made of elementary
plaquettes.

In our case these surfaces would be made from the variable loops,
glued together side by side, in the same way, as elementary plaquettes
of the Wilson expansion. Our surfaces are breathing, we have some
"matter" on the world sheet of string. However, we do not want this
matter to show up in the physical spectrum: we adjust the parameters
so, that these scalar fields stay heavy.

Coming back to the continuum limit, let us note, that the reason for
leaving only the Yang-Mills term here repeats the
original physical motivation for QCD to be the theory of hadronic interaction:
regardless of what could be the quark and gluon constituents,
the effective theory
above Grand Unification scales must be QCD as the only renormalizable theory,
with the effective constant, running from something large at Planck scales to
correct coupling at prehadronic scales. In our model, we take a
 particular choice
of gluon constituents, which by universality and renormalizability must induce
correct QCD.

The scalar particle mass $ m $ would now serve as the ultraviolet cutoff for
QCD, as it is the scale when the gluon interaction becomes nonlocal. Repeating
the standard arguments of asymptotic freedom we would find the RG relation
 between the bare coupling and the glueball mass $ \mu $
\begin{equation}
\ln \frac{m^2}{\mu^2}  \rightarrow  \frac{48 \pi^2}{11Ng_0^2}
\label{AsymptoticFreedom}
\end{equation}
Comparing these two relations we find the scaling law:
\begin{equation}
\mu^2 \rightarrow  ( m_0^2 - m_c^2)^{\frac{23}{22}}
\label{ScalingLaw}
\end{equation}
However, this relation was derived without taking into account the
feedback of hard gluons, so that one could use it only as a hint. At
the moment it seems, that there is no universal scaling index. The
real issue is, of course, whether the model would escape the Gaussian
fixed point at large scale. This would mean , that we are dealing with
gauge theory.

To summarize, this line of argument involved integration first over the scalar
 field
to induce effective Yang-Mills theory, and then over gauge field to derive the
RG law for the glueball spectrum. This is not the best way to solve our model,
which was designed for the opposite integration order.

\section{Effective Interaction for Master Field}

Let us now integrate our theory in the opposite order, i.e. over the gauge
fields first. This is possible in absence of direct gauge coupling in the bare
Lagrangean. All we need , is the one link integral, which was computed first by
Itzykson and Zuber\cite{ItZu} (and was probably  known before to
mathematicians \cite{HCh})
\begin{equation}
I(\phi,\chi) = \int D U \exp \left(
N \, \tr \phi U \chi U^{\dagger}
\right) \propto \frac{\det_{ij} \exp(N \phi_i \chi_j )}{\Delta(\phi)
\Delta(\chi)}
\label{IZ}
\end{equation}
where
\begin{equation}
\Delta(\phi) = \prod_{i<j} (\phi_i-\phi_j)
\label{Vandermond}
\end{equation}
is the Vandermonde determinant.

Using eq. (/ref{IZ}) we find from eq.(\ref{Action}) the following
partition function of our theory in terms of only eigenvalues
$\phi_{i}$ of the original scalar field:
\begin{equation}
Z= \int \prod_{x} [d^{N}\phi(x) \exp(-N/2 \sum_{i} \phi(x)^{2})
\Delta^{2}(\phi(x))] \prod_{<xy>}
\frac{\det_{ij} \exp(N \phi_{i}(x) \phi_{j}(y)}{\Delta(\phi(x))
\Delta(\phi(y))}
\label{partfun}
\end{equation}
where $<xy>$ denote the neighbouring vertices $x$ and $y$ on the
lattice.

Note thatabovementioned naive consideration of the continuum model we
would obtain only the diagonal term $exp(N\phi_i \phi_j)$ from the
determinant, with all the unwanted phenomena like the collaps of the
eigenvalues, whereas in the correct lattice version these problems are
absent.

Should we work with the physical $ N=3 $ theory, this would be the
explicit result for the
effective interaction of the eigenvalues.
However, in the large $N$ limit, it is only half way to the result.

The problem is obvious: there is the sign-alternating sum of $ N! $ terms, each
of the order of $ \exp(N^2) $. They must cancel each other to some extent, to
compensate the vanishing denominators at coinciding eigenvalues, since the
integral of a positive finite function
 over a unitary group is always positive and
finite.  Instead of tracing these cancellations we use a different
representation for the whole integral
\begin{equation}
I(\phi,\chi) \propto \int d^N x d^N y
\frac{\Delta(x)\Delta(y)}{\prod_{a,i}(x_i-\phi_a)(y_i-\chi_a)}\exp \left(
N \sum_i x_i y_i
\right)
\label{XYintegral}
\end{equation}
where the integration contour over $ x (y)$ encircles the spectrum of
eigenvalues $ \phi_a (\chi_a) $.
The proof is straightforward: one should compute the residues at the poles,
which produce  squares of Vandermonde determinants in denominator, combining
with the same determinants in the numerator, after which the sign-alternating
sum over all possible pole terms produces the determinant in the numerator.

But this is the same as the two matrix integral
\begin{equation}
\int d^N x d^N y \Delta(x)\Delta(y) \exp \left(
N \sum_i x_i y_i - V_1(x_i) -V_2(y_i)
\right)
\label{TwoMatrix}
\end{equation}
with singular potentials
\begin{equation}
V_1(x) = \frac{1}{N} \sum_a \ln(x-\phi_a) \\;\;
V_2(y) = \frac{1}{N} \sum_a \ln(y-\chi_a)
\label{Potentials}
\end{equation}

This integral can be computed at large $N$ by means of the orthogonal
polynomial technique\cite{Mehta}. For the reader's convenience we reproduce
this
computation in Appendix. We assume here certain analytic properties of the
potential, which we later find selfconsistent.

The resulting equations of the two matrix models for the relevant case of $
\chi= \phi, V_1=V_2 =V$ read
\begin{equation}
\ln I(\phi,\phi) = N \ln h_1 +N^2 \int_0^1 dt(1-t)\ln f(t)
\label{FreeEnergy}
\end{equation}
\begin{equation}
h_1 =\oint \frac{dx}{2 \pi \imath} \oint \frac{dy}{2 \pi \imath}\exp( N x y - N
V(x) - N V(y))
\end{equation}
\begin{equation}
f(t) = -t + \oint \frac{dz}{2\pi \imath} \frac{V'(q(z,t))}{z^2}
\label{FT}
\end{equation}
\begin{equation}
q(y,t) = \frac{1}{y} + \oint \frac{dz}{2\pi \imath}
\frac{V'(q(z,t))}{z\left(1-y z f(t)\right)}
\end{equation}
where the $ z $ integration goes counterclockwise  surrounding all the
singularities of the function $ V'(q(z,t)) $.

The last equation represents the integral equation for $ q(z,t) $ at fixed $t$,
 the second equation expresses $ t $ in terms of $ f $ , while the first one
yields the answer for the large $ N $ limit of the Itzykson-Zuber integral.
In the classical field equation of the next Section we shall need only the
derivatives which we compute in the Appendix with the result
\begin{equation}
\frac{1}{N}\frac{\partial \ln I(\phi,\chi)}{\partial \phi_a} =  \int_0^1 dt
\oint \frac{dz}{2\pi \imath z} \frac{1}{q(z,t)-\phi_a}
\label{Derivative}
\end{equation}

In the usual case of the polynomial potentials
the solution for $ q(z,t) $  has the
form of a Laurent expansion
\begin{equation}
q(z,t) = \frac{1}{z} + \sum_{n} q_n(t)z^n
\end{equation}
which terminates at some
integrals reduced to the residue at $ z=0 $ which produces a set of algebraic
relations for the parameters $ q_n(t),f(t) $.

The critical behaviour arises here by the standard mechanism, recently
investigated in great detail in the matrix models of 2D Quantum
Gravity\cite{BrKa},
\cite{GrMi}, \cite{DoSh}.
 Namely, the parameters must be adjusted so that the poles of the integrand in
(\ref{Derivative}) pinch the integration contour at the symmetry points $ z=
z^{-1} $. One could study, the requirement
that the linear terms vanish in $
q(z,t)$ at $ z^2=t=1$.
\begin{equation}
\frac{\partial q(z,1)}{\partial z} =0 \\;\; z^2 =1.
\label{CriticalPoint}
\end{equation}
In the vicinity of this point there would be powerlike singularities, the
square root in the absence
 of further parameter adjustment. The structure of these
singularities depends on the potential, which could be adjusted to produce any
rational singularities.

In our case the potential is not a polynomial. In general, we expect branch
point singularities  corresponding to  finite support of eigenvalues. In
this case the above integral equations should be investigated in the vicinity
of
these singularities together with the classical field equation for the density
 of
these eigenvalues, which we derive in the next Section.

Let us stress once again that the above formulas imply  analyticity  of the
potential at the origin. One cannot apply them to the truncated expansions of
the potential $ V'(q) $ in  negative powers of $ q $. The expansion in inverse
powers of $ q $ corresponds to expansion in inverse powers of $ m_0$. This
expansion is studied in the next Section.

\section{Strong Coupling Expansion}

We expect a phase transition in our model at a
 certain critical value $ m_c $ of
the bare mass $ m_0 $, and we are interested in the strong coupling phase, when
$ m_0 \searrow m_c $. The above general solution applies to the opposite phase
$ m_0 < m_c $, which will become clear below. In this Section we are going to
study the strong coupling phase. Unfortunately, we could not find an explicit
analytic solution in this phase, but we have
found the strong coupling expansion.

Let us rescale $ \phi \rightarrow \frac{\phi}{m_0} $, $ \chi \rightarrow
\frac{\chi}{m_0} $ and expand our potentials in $ \frac{1}{m_0} $
\begin{equation}
V_1(x) \rightarrow \ln x - \sum_{n>0} \frac{t_n}{n m_0^n x^n} \\; \;
V_2(x) \rightarrow \ln y - \sum_{n>0} \frac{s_n}{n m_0^n y^n}
\end{equation}
where
\begin{equation}
t_n = \int d\phi \rho(\phi) \phi^n \\;\;
s_n = \int d\chi \rho(\chi) \chi^n
\end{equation}
are the moments of the distributions of eigenvalues. We shall assume that odd
moments vanish, which corresponds to unbroken parity of our model.

As it turns out, the above general solution of the two matrix model does not
work in this expansion. In any order we find
\begin{equation}
q(z) = \frac{1}{z} \\;\; f(t) = 1-t \\;\;
\frac{\partial \ln I(\phi,\chi)}{\partial \phi_a} =0
\end{equation}
as exact solution.

But we know that there should be $ \frac{1}{m_0} $ corrections. Straightforward
expansion in powers of $ N \, \tr \phi U \chi U^{\dagger} $ in the original
definition and $ U(N) $ group integration using Schwinger-Dyson equations
yields, at infinite $ N $,
\begin{equation}
\frac{1}{N}\frac{\partial \ln I(\phi,\chi)}{\partial \phi_a} = \Gamma_2 \phi_a
+ \Gamma_4 \phi_a^3 +\Gamma_6 \phi_a^5 + O\left(m_0^{-15}\right)
\end{equation}
where
\begin{eqnarray}
\Gamma_2  &=&
 {{s_{2}}\over {{{m_{0}}^3}}} +
  {{3\,{{s_{2}}^2}\,t_{2} - 2\,s_{4}\,t_{2}}\over
    {{{m_{0}}^7}}} \\ \nonumber
    &+&
 {{27\,{{s_{2}}^3}\,{{t_{2}}^2} -
      30\,s_{2}\,s_{4}\,{{t_{2}}^2} + 7\,s_{6}\,{{t_{2}}^2}
       - 10\,{{s_{2}}^3}\,t_{4} +
      10\,s_{2}\,s_{4}\,t_{4} - 2\,s_{6}\,t_{4}}\over
    {{{m_{0}}^{11}}}}\\ \nonumber
\Gamma_4  &=&
  {{-2\,{{s_{2}}^2} + s_{4}}\over {{{m_{0}}^7}}} +
  {{-20\,{{s_{2}}^3}\,t_{2} +
      20\,s_{2}\,s_{4}\,t_{2} - 4\,s_{6}\,t_{2}}\over
    {{{m_{0}}^{11}}}} \\ \nonumber
\Gamma_6 &=&
   {{7\,{{s_{2}}^3} - 6\,s_{2}\,s_{4} + s_{6}}\over
   {{{m_{0}}^{11}}}}
\label{Gammas}
\end{eqnarray}

The paradox is resolved by neglected singularities at small $ t$ in our formal
solution. With a regular potential there are no singularities at small $ t $ so
that the weak coupling phase is OK, but in the strong coupling phase the
contributions from small $ t \sim \frac{1}{N} $ in the above integrals
dominate.

To see this, let us note that within the given order $L$ of the
 $ \frac{1}{m_0} $
expansion one could write
\begin{equation}
\exp\left(N(xy - V_1(x) -V_2(y)) \right) =
\frac{\exp\left(Nxy\right)}{x^N R_1(x)y^N R_2(y)}
\end{equation}
where
\begin{equation}
R_1(x) = 1 + \frac{N t_2}{2 m_0^2 x^2} + \dots \\;\;
R_2(y) = 1 + \frac{N s_2}{2 m_0^2 y^2} + \dots
\end{equation}
are the $L$-degree polynomials in the inverse argument.
 The fact that there are
higher  powers of $N$ in these polynomials does not bother us at the moment,
since we are expanding in $ \frac{1}{m_0} $ at fixed $N$, and we know that
higher order terms  would cancel in the final equations.

This observation allows us to explicitely construct all the orthogonal
polynomials with large order
\begin{equation}
P_l(x) = x^{l-1} R_1(x) \\;\;
Q_l(y) = y^{l-1} R_2(y) \\;\; l > L
\end{equation}
Indeed, these are polynomials with proper normalization at infinity, and the
integrals in orthogonality relation
\begin{equation}
\left\langle P_n(x) Q_m(y) \right\rangle \stackrel{n,m > L}{\Longrightarrow}
\oint \frac{dx}{2 \pi \imath} \oint \frac{dy}{2 \pi \imath}\exp( N x y )
x^{n-N-1} y^{m-N-1} = \delta_{n m} h_n \\;\;
h_n =\frac{N^{N-n}}{(N-n)!}
\end{equation}
correspond to $ f(t) = 1-t $ at $ N t> L$. One can readily check that these
polynomials $ P_l(x) $ are orthogonal to all powers $ x^k \, ,k < l-1 $. The $
y $ integral reduces to residue at zero, after which the $ x $ integral reduces
to residue at infinity, which vanish. The same is true for $ Q_l(y) $, with
the obvious change of variables.

Therefore, the derivative of our integral reduces to the  sum of $L$ terms
\begin{equation}
\frac{\partial \ln I(\phi,\chi)}{\partial \phi_a} =
\sum_{n=1}^L \left\langle \frac{P_n(x) Q_{n}(y)}{(x-\phi_a)h_n} \right\rangle
\end{equation}
$ \frac{1}{m_0} $ expansion of these terms can be explicitely computed using
recurrense equations for the polynomials,
working down to the zeroth and first order
polynomials which are trivial.
We have checked that the correct terms of $ \frac{1}{m_0}
$ expansion are reproduced this way.

One may wonder why the lowest order integrals , such as $ h_1$, cannot be
computed by the saddle point approximation at large $ N $. The answer is
interesting. Within the $ \frac{1}{m_0} $ expansion there are multiple
degenerate saddle points, so that one has to average over them, which leads to
cancellations. One could also reproduce correct answers this way with some
effort.
The resume is that $ \frac{1}{m_0} $ expansion can be carried through to high
order, which could be used for numerical computations.

\section{Classical Field Equation}

We are approaching the most surprising part of our theory, namely, the
classical
field equation. The fact is, the integration over link variables $ U_{\mu}(x) $
eliminates the angular variables of the scalar field, leaving us with only
 $ N
$ eigenvalues at each cite of the lattice. Some 15 years ago, without all the
frustrating experience of  large $ N $ QCD, one would not hesitate to apply
the WKB approximation to the remaining functional integral over the
eigenvalues. But now, the sophisticated reader might wonder where are the
planar graphs.

The answer is, we have already  summed them up!
 The angular variables $\Omega(x) $ of
the scalar field $ \Phi(x) = \Omega^{-1}(x)\phi(x)\Omega(x) $ which we
integrated at each site as the gauge part of the link integration, $ U_{\mu}(x)
\rightarrow  \Omega^{-1}(x)U_{\mu}(x)\Omega(x+ \mu) $ represent the majority of
scalar degrees of freedom, responsible for the planar graphs.

For example, take the box diagram, corresponding to the one plaquette loop of
the scalar particle. In conventional approach, one would expand
in the hopping term
at the original Action, and get the integral
\begin{equation}
\int DU D\Phi \exp \left( -\sum_x N \, \tr \left( m_0^2 \Phi^2(x)
\right) \right)
\prod_{i=1}^4 \tr\left(\Phi(i) U(i) \Phi(i+1) U^{\dagger}(i)\right)
\;;\; \Phi(5) \equiv \Phi(1)
\end{equation}
Integrating  over $ \Phi $ first, one would get the usual  Feynman graphs in
an external gauge field,
 which then would be integrated out in the strong coupling
approximation ( this is because we have not yet sumed up the scalar loops to
 find
finite gauge coupling ). Integrating the box diagram in  reverse order, we
get a product of traces
\begin{equation}
N^4 \prod_{i=1}^4\int DU(i) \tr \left(\Phi(i) U(i) \Phi(i+1)
U^{\dagger}(i)\right) = \prod_{i=1}^4 (\tr\Phi(i))^2
\end{equation}
which yields zero since our scalar field is traceless. One could check that the
same result comes about from the Feynman graphs of the scalar field. In a less
trivial example of the double box, the integrations of
\begin{equation}
 N^8 \prod_{i=1}^4\int DU(i) \left(\tr \left(\Phi(i) U(i) \Phi(i+1)
U^{\dagger}(i)\right)\right)^2
\end{equation}
already yield a nonzero result.
As in the previous example, the integration over
link  matrices yields the product of the traces in all vertices, but this time
there would be traces of $ \Phi^2(i) $. At large $ N $ these traces can be
computed in the WKB approximation for the eigenvalues $ \phi_a $,
 which for the
Gaussian field corresponds to the famous semicircle distribution
\begin{equation}
\rho(\phi) \propto \sqrt{2- m_0^2 \phi^2 }
\end{equation}

Analyzing such examples, we come to the conclusion that link integrations take
care of all the relevant quantum fluctuations at large $ N $, the scalar field
as well as the gluon field. These are quite nontrivial integrals.
They can be regarded as sums over $ N! $ one loop WKB integrals around
nontrivial ``gauge vacua ", each integral producing two Vandermonde
determinants in the denominator. All the ``higher loop corrections "
identically
vanish, as a result of deep mathematical theorems about group manifolds. This
means that we have summed over infinite number of ``instantons" of our lattice
theory, corresponding to one of each of $ N! $ classical solutions at each
link, plus the harmonic quantum fluctuations.

Let us now write down the classical equation for the eigenvalues of the scalar
field, assuming that it is
spatially uniform to fulfill  translational invariance of
the vacuum. Eliminating the angular variables, we get the square of the
Vandermonde determinant at each vertex. The resulting saddle point equation for
the constant master field reads
\begin{equation}
2 \sum_{b \neq a} \frac{1}{\phi_a - \phi_b} = 2 N m_0^2 \phi_a -
2 d\left[\frac{\partial \ln I(\phi,\chi)}{\partial \phi_a}\right]_{\chi=\phi}
\end{equation}

Let us consider the weak coupling phase first.
Using formulas of the previous Section we find here at $ N = \infty $:
\begin{equation}
\frac{1}{2}\left(V'(\phi^{+}) + V'(\phi^{-})\right) = m_0^2 \phi - d \int_0^1
dt \oint \frac{dz}{ 2\pi \imath z}\frac{1}{q - \phi} \;;\; q \equiv q(z,t)
\label{MasterField}
\end{equation}
where $ \phi^{\pm} $ correspond to the opposite sides of the branch cut in the
complex plane, corresponding to the spectrum of the eigenvalues. The equation
holds only at this branch cut.

Let us now introduce the analytic function $ F(\phi^2) = \phi V'(\phi) $ (we
assume that the parity transformation $ \phi \rightarrow - \phi $ of our theory
is not spontaneously broken). The integral in (\ref{MasterField}) is, in fact,
odd function of $ \phi $ as $ q $ is an odd function of $ z $ for an odd
potential, which allows us to rewrite the
equation for F in manifestly symmetric
form
\begin{equation}
\frac{1}{2}\left(F(\xi^{+}) + F(\xi^{-})\right) = m_0^2\xi + d - d \int_0^1 dt
\oint \frac{dz}{ 2\pi \imath z}\frac{q^2}{q^2 - \xi}
\label{Fequation}
\end{equation}

This is  a boundary problem for the analytic function $ F(\xi) $, with the
conditions  $ F(\infty ) = 1 $ and $ F(0) = 0 $. The following Ansatz can be
written down
\begin{equation}
F(\xi) = m_0^2 (\xi - S(\xi) ) + d - d \int_0^1 dt \oint \frac{dz}{ 2\pi \imath
z}\frac{q^2(S(q^2)-S(\xi))}{S(q^2)(q^2 - \xi)} \;;\; S(\eta) \equiv
\sqrt{(\eta-a^2)(\eta - b^2)}
\end{equation}
This Ansatz satisfies the above equation, as $ S(\xi^{+}) = - S(\xi^{-}) $ at
the cut between $ a^2  $ and $ b^2 $. The fictitious singularities at $ \xi
=q^2 $ cancel between numerator and denominator of the integral. As for the
conditions at $ 0 $ and $ \infty $, they yield the following equations
\begin{equation}
m_0^2 = \frac{2}{a^2 + b^2} \left(1 -d + d \int_0^1 dt \oint \frac{dz}{ 2\pi
\imath z}\frac{q^2}{S(q^2)} \right)
\label{Mequation}
\end{equation}
\begin{equation}
0 = a b \left(-m_0^2 +d \int_0^1 dt \oint \frac{dz}{ 2\pi \imath
z}\frac{1}{S(q^2)} \right)
\label{Requation}
\end{equation}

The trivial solution, with $ a=0$ or $b=0$ does not apply, since we assumed
regular behaviour of $ F(\xi) $ at the origin. Hence, we should choose
finite solutions for $ a,b $. This solution for the potential should now be
substituted back into equations for $ q(z,t), f(t) $ of the previous Section.
At present we do not know how to solve these equations, this seems to be a
serious mathematical problem.

As for the strong coupling phase, here we note that within the $ \frac{1}{m_0}
$ expansion the master field equation coinsides with that of the planar limit
of the one matrix model with $ \beta =  N d $ and effective potential
$ U_{eff}(\phi) $ such that
\begin{equation}
U'_{eff}(\phi) = \left(\frac{m_0^2}{d}- \Gamma_2\right)\phi
-\Gamma_4 \phi^3
-\Gamma_6 \phi^5
\end{equation}
Therefore, the $ \frac{1}{m_0} $ expansion is generated by the planar graphs of
this model. The  well known equations of the orthogonal polynomial the
solution of
spherical limit of the one matrix model can be applied now
\begin{equation}
\frac{t}{d} =  \oint \frac{dz}{2 \pi
\imath}U_{eff}'\left(z+\frac{R(t)}{z}\right)=
\left(\frac{m_0^2}{d}- \Gamma_2\right)R(t) -
\sum_{n>2}\frac{(2n)!}{(n!)^2}\Gamma_{2n} R(t)^n
\end{equation}
\begin{equation}
t_{2n} = s_{2n} = \frac{(2n)!}{(n!)^2} \int_0^1 dt \left(m_0^2R(t)\right)^n
\end{equation}
\begin{equation}
\rho(\phi) = \frac{1}{2 \pi} \int_0^1 dt \; \frac{\theta\left(4 R(t) -
\phi^2\right)}{\sqrt{4 R(t) - \phi^2}}
\end{equation}
The first relation provides an implicit equation for the function $ R(t) $,
whereas the second equation yields the selfconsistency relations for the
moments, involved in coefficients $ \Gamma_k $ defined in (\ref{Gammas}).

These equations could be expanded in $ \frac{1}{m_0^2} $
\begin{equation}
R(t) = \frac{t}{2m_0^2} + \dots \\;\; s_{2n} = \frac{(2n-1)!!}{(n+1)!} +\dots
\\;\;
\rho\left(\frac{x}{m_0}\right) = \frac{m_0}{2\pi}\sqrt{2-x^2}+\dots
\end{equation}

The eigenvalues are distributed from $ -b $ to $ b $  where $ b^2 =4 \max R(t)
\, , 0<t<1 $ within the $ \frac{1}{m_0} $ expansion. The phase transition to
the above weak coupling phase would occur when the generating function
\begin{equation}
F(\xi) = \frac{1}{2 \pi}  \int_0^1 dt\;\frac{1}{\sqrt{\xi-4R(t)}}
\end{equation}
develops a  singularity at $ \xi =0$. This singularity must come out from the
second sheet, as there are no singularities at the first sheet within the $
\frac{1}{m_0} $ expansion. The density $ \rho(\phi) $ is a positive function,
with a maximum at $ \phi = 0 $, and therefore
we expect $ \rho(0) $ to grow as some
power of $ m_0^2 - m_c^2$.
At the critical point the density diverges, after which we expect the gap from
$ -a $ to $a $ to arise in the eigenvalue distribution, in accordance
 with the weak
coupling solution.

In order to test this scenario one could expand the logarithmic derivative   $
\frac{\partial \ln \rho(0)}{\partial  m_0} $ in $ \frac{1}{m_0^2} $ and Pad\'e
extrapolate to find the pole. The analytic computation of this series to
high order is tractable, but exceeds our present scope. It seems to us that so
far the basic physical properties of this remarkable model are  more important
then its numerical solution.

\section{Wave Equation}

Let us now discuss the excitations in our vacuum. In the harmonic
approximation,
corresponding to the next term of the large $ N $ expansion, we find in
effective master field Action\footnote{One may check that the $
\delta_{a b} $ terms all cancel.}
\begin{equation}
S_2 = \sum_x \sum_{a \neq b}
        \left( \frac{1}{(\phi_a-\phi_b)^2}
           -d \sigma_{a b}
        \right)
\delta \phi_a(x) \delta \phi_b(x) - \frac{1}{2} \eta_{a b} \delta \phi_a(x)
\sum_{\mu=-d}^{d}
\delta \phi_b(x+ \mu)
\end{equation}
where the matrix of second derivatives is computed in the Appendix
\begin{eqnarray}
& &\sigma_{a b} = \oint \frac{dz_1}{2\pi \imath} \oint \frac{dz_2}{2\pi \imath}
\frac{1}{(z_1-z_2)^2 (q(z_1,1)-\phi_a)(q(z_2,1)-\phi_b)} \\ \nonumber
& \,&
\eta_{a b} =\oint \frac{dz_1}{2\pi \imath} \oint \frac{dz_2}{2\pi \imath}
\frac{f(1)}{(1-f(1)z_1z_2)^2 (q(z_1,1)-\phi_a)(q(z_2,1)-\phi_b)}
\end{eqnarray}

As it is argued in the
Appendix, the pole at $ z_1 = z_2 $ should be treated in the
principal value sense, while the pole at $ f(1) z_1 z_2 =1 $ lies outside the
integration contours, which are little circles surrounding remaining
singularities of the integrand.

Note that the next corrections for the vacuum field do not display themselves
 in
the leading order for the second variation matrix; it can be computed at $ N =
\infty $. The factor of $ N $ cancels when the field variation is rescaled $
\delta \phi = \frac{\psi(\phi)}{N} $ and all the  sums are converted to the
integrals in the local limit
\begin{eqnarray}
S_2 &=& \int d^d x {\cal L} \\ \nonumber
{\cal L } &=&
        \int  d\phi \rho(\phi) \int  d\phi'  \rho(\phi') \psi(\phi)
        \left(
         B(\phi,\phi')\psi(\phi') +
        C(\phi,\phi') \partial^2 \psi(\phi')
        \right)
\label{S2Continuum}
\end{eqnarray}
where
\begin{eqnarray}
C &=& -\frac{f(1)}{2} \oint \oint \frac{dz_1dz_2}{(2\pi \imath)^2}
\frac{1}{(1-f(1)z_1z_2)^2 (q(z_1,1)-\phi)(q(z_2,1)-\phi')} \\ \nonumber
B &=& \frac{1}{(\phi-\phi')2} + d C - d \oint \oint \frac{dz_1dz_2}{(2\pi
 \imath)^2}
\frac{1}{(z_1-z_2)^2(q(z_1,1)-\phi)(q(z_2,1)-\phi')}
\end{eqnarray}
and double poles are to be treated in the principal value sence.

The particle masses $ M_i $ are to be found from
the wave equation
\begin{equation}
\int d\phi' \rho(\phi') \left(C(\phi,\phi') M_i^2 -
B(\phi,\phi')\right) \psi(\phi') = 0
\label{WaveEquation}
\end{equation}
which in general has a discrete spectrum, for the finite support of
eigenvalues. Clearly, the bare mass $ m_0 $  should be adjusted to make the
physical mass scale small in the lattice units we are using. Presumably, the
scaling law of a type (\ref{ScalingLaw})
would come out of this equation, or from the
similar equation of the strong coupling phase. The latter equation is yet to be
derived, which is a challenge to  large $ N $ experts.

The serious problem with the weak coupling solution under
consideration is that apparently $ f(1) =0 $ so that there is no
hopping term in effective Action, and therefore no mass spectrum.

The arguments for vanishing $ f(1)$ are very simple. As discussed in
the Section 4, the orthogonal polynomials of high enough order can be
explicitely constructed. In particular, for $ L > N $
\begin{equation}
P_L(x) = x^{L-N-1} \prod_{i=1}^{N} (x-x_i) \\; \;
Q_L(y) = y^{L-N-1} \prod_{i=1}^{N} (y-y_i)
\end{equation}
so that
\begin{equation}
h_L = \frac{N^{N-L}}{\Gamma(N+1-L)} \\; \; f \left( \frac{L}{N} \right) = 1-
L/N
\end{equation}
which provides us with boundary value $ f(1) =0 $.

\section{Discussion}

So, is this the long-anticipated exact solution of large $ N $ QCD? Not quite,
but we are getting closer to it. It remains to be proven that this theory
correctly induces QCD, which would require either an
 exact solution in the strong
coupling phase, or  numerical simulations. In our opinion, both are highly
desirable.

As for numerical simulations, these would be especially simple at $ N=2$ where
there are only two eigenvalues of  opposite sign $ \pm \phi $, and the
functional  integral reduces to
\begin{equation}
\prod_{x}\int_0^{\infty} d \phi(x) \phi(x)^2 \exp\left(- 2 m_0^2
\phi(x)^2\right) \prod_{<x y>} \frac{\sinh(4 \phi(x) \phi(y))}{4 \phi(x)
\phi(y)}
\label{Neq2}
\end{equation}
We do not see any reason why our model should fail to induce QCD at $ N=2 $ as
well as it does at $ N= \infty $. Clearly, the WKB approximation is not
supposed to work so well here, but at least we could expect the vacuum average
of $ \phi $.
The restriction $ \phi >0 $ could be taken into account by the change of
variables
$ \phi = \exp( \lambda ) $ where $ \lambda $ varies from $ - \infty $ to $ +
\infty $. Still, the potential for the $ \lambda $ field remains essentially
nonlinear, so that it could avoid the zero charge problem.
The model looks too simple, and maybe some flaw in it
would be found, but it is worth
a try. At least we would get some insight and see what could go wrong.

Then one could  try the $ SU(3) $ model. What are the phases of this model?
How does it compare with the
usual QCD? Maybe the scaling laws would be easier to
obtain numerically
than the exponential laws of asymptotic freedom? The scaling
indices in, say, the $ 3D $ Ising model are reproduced by  state of the art
lattice simulations with several digits, whereas  asymptotic freedom
has
never been correctly reproduced, to the
great embarrassment of the lattice gauge
community. So, why not try our model instead? It does not look much harder than
the Ising model.

As for exact solution, the part of the theory which we miss badly is the strong
coupling representation of the Itzykson-Zuber determinant in the large
$N$ limit. We feel that such a solution is just around the corner, but all
efforts to find it from the two matrix model representation have failed.

Still, we could approach the critical point from the weak coupling side, which
we did in the previous Section. There are explicit nonlinear integral equations
for the eigenvalue density and the wave equation for the glueball spectrum to
solve, which is another interesting numerical problem.

\section{Acknowledgements}

We are grateful to E.Brezin, M.Douglas, D.Gross,
Al.Zamolodchikov,
I.K.Kostov, I.Klebanov,  A.Polyakov and S. Shenker for fruitful
 discussions and
important remarks.  AAM was  partially supported by the National Science
Foundation under contract PHYS-90-21984.

\appendix

\section{Itzykson-Zuber determinant via Orthogonal Polynomials
in Two Matrix Model}

Let us reproduce and slightly generalize  Mehta's solution of the two matrix
model by means of orthogonal polynomials. The basic idea is to introduce
orthogonal polynomials $ P_n(x) $, $  Q_m(y) $ with properties
\begin{equation}
\left\langle P_n(x) Q_m(y) \right\rangle  = h_n \delta_{n m} \\;\;
\left\langle F(x,y)\right\rangle \equiv \oint \frac{dx}{2 \pi \imath} \oint
\frac{dy}{2 \pi \imath}F(x,y)\exp( N x y - N V_1(x) - N V_2(y))
\label{Orthogonality}
\end{equation}
The integration contour surrounds the singularities of the measure, which in
our case are simple poles at $ x = \phi_a $ and $ y = \chi_b $. The
normalization of polynomials is such that they tend to $ P_n(x) \rightarrow
x^{n-1} \, , Q_m(y) \rightarrow y^{m-1} $ at infinity. Note that we shifted the
index by one unit with respect to traditional definition - this simplifies
formulas below. The Vandermonde determinants can be expressed as determinants
of these polynomials
\begin{equation}
\Delta(x) = \det_{i,j} P_j(x_i) = \epsilon_{\{n\}}\prod_{i} P_{n_i}(x_i)
\end{equation}
\begin{equation}
\Delta(y) = \det_{i,j} Q_j(x_i) = \epsilon_{\{m\}}\prod_{i} Q_{m_i}(y_i)
\end{equation}
where $ \epsilon_{\{n\}} $ denotes the unit antisymmetric tensor with $ N $
indices $ n_1,\dots n_N $.

With these substitutions our original integral reduces to
\begin{equation}
I \propto \epsilon_{\{n\}} \epsilon_{\{m\}}\prod_{i} \left\langle P_{n_i}(x_i)
Q_{m_i}(y_i) \right\rangle = N! \prod_{n=1}^N h_n
\label{Hproduct}
\end{equation}
A convenient set of parameters, with smooth limit at large $ N $ is given by
the coefficients $ q_l(t) \, , p_l(t)  $ of the expansion ( $  t \equiv
\frac{n}{N} $ )
\begin{equation}
x P_n(x) = P_{n+1}(x) + \sum_l p_l(t) P_{n-l}(x) \\;\;
y Q_n(y) = Q_{n+1}(y) + \sum_l q_l(t) Q_{n-l}(y)
\label{PQexpansion}
\end{equation}
which for finite $ N $ terminates at $ l= N $ but in the large $ N $ limit
becomes an infinite expansion. The equations for these parameters follow from
identities
\begin{equation}
0 = \left\langle P_{n-l}'(x) Q_{n}(y) \right\rangle
\end{equation}
Integrating by parts in $ x $ we find
\begin{equation}
\left\langle P_{n-l}(x) (V_1'(x) - y ) Q_{n}(y) \right\rangle = 0
\end{equation}
Now, using the expansion (\ref{PQexpansion}) together with orthogonality
condition (\ref{Orthogonality}) we arrive at the relation
\begin{equation}
q_{l}(t) = h_{n-l}^{-1}\left\langle P_{n-l}(x)V_1'(x)Q_{n}(y) \right\rangle
\end{equation}
and the similar relation for $ p_{l}(t) $
\begin{equation}
p_{l}(t) = h_{n-l}^{-1}\left\langle Q_{n-l}(y)V_2'(y)P_{n}(x) \right\rangle
\end{equation}
There is one more relation, for the $ h_n $ coefficients. It follows from the
identity
\begin{equation}
0 = \left\langle \left(P_{n+1}'(x)-n P_n(x) \right) Q_n(y) \right\rangle
\end{equation}
after integration by parts of the first term
\begin{equation}
0 = \left\langle \left(P_{n+1}(x) N(V_1'(x)-y) - n P_n(x) \right) Q_n(y)
\right\rangle
\end{equation}
and using orthogonality relation
\begin{equation}
h_{n+1}= - \frac{n}{N} h_n + \left\langle P_{n+1}(x) V_1'(x) Q_n(y)
\right\rangle
\end{equation}

At this point it is convenient to introduce the ratio
\begin{equation}
f(t) = \frac{h_{n+1}}{h_n} \\;\; t \equiv \frac{n}{N}
\end{equation}
and the normalized $ P\,,Q $ polynomials
\begin{equation}
\tilde{Q}_{n}(y) = \frac{Q_n(y)}{h_n} \\; \;
\tilde{P}_{n}(x) = \frac{P_n(x)}{h_n}
\end{equation}
\begin{equation}
\left\langle P_n(x) \tilde{Q}_{m}(y) \right\rangle =
\left\langle Q_m(y) \tilde{P}_{n}(x) \right\rangle = \delta_{n m}
\end{equation}
and rewrite  above equations as follows
\begin{equation}
\ln I = \ln N!  + N \ln h_1  + \sum_{k=1}^{N-1} (N-k) \ln f\left(\frac{k}{N}
\right)
\end{equation}
\begin{equation}
f(t) = -t + \left\langle P_{n+1}(x) V_1'(x) \tilde{Q}_{n}(y) \right\rangle
\end{equation}
\begin{equation}
q_l(t) = \left\langle P_{n-l}(x) V_1'(x) \tilde{Q}_{n}(y) \right\rangle
\prod_{k=1}^l f\left(t-\frac{k}{N}\right)
\label{Qequation}
\end{equation}
\begin{equation}
p_l(t) = \left\langle Q_{n-l}(y) V_2'(y) \tilde{P}_{n}(x) \right\rangle
\prod_{k=1}^l f\left(t-\frac{k}{N}\right)
\end{equation}

As for the derivative of the free energy, it reduces to the following
\begin{equation}
\frac{\partial \ln I(\phi,\chi)}{\partial \phi_a} =
\sum_n \left\langle \frac{P_n(x) \tilde{Q}_{n}(y)}{x-\phi_a} \right\rangle
\end{equation}

Now, assuming existence of the large $ N $ limit of $ f $ we arrive at the
asymptotic formula for our integral
\begin{equation}
\frac{\ln I }{N^2} \rightarrow \int_0^1 dt (1-t) \ln f(t)
\end{equation}

The products in above equations converge to
\begin{equation}
\prod_{k=1}^l f\left(t-\frac{k}{N}\right) \rightarrow f^l(t)\left(1 -
\frac{l(l+1)}{2 N} \ln \frac{f'(t)}{f(t)} \right) + \dots \rightarrow f^l(t)
\end{equation}

As for the normalized matrix elements, those could be computed at large $ N $
in the WKB approximation, using the shift operator
\begin{equation}
z = \exp \left(- \frac{d}{d n} \right)
\end{equation}
and neglecting its commutators with $ t $
\begin{equation}
[ z \, , t ] = -\frac{1}{N} \rightarrow 0
\end{equation}
The recurrent equations for our polynomials can be rewritten as operator
relations
\begin{equation}
y Q_{n-k}(y) = q(z,t) z^k Q_n(y) \\;\;
x P_{n-k}(x) = p(z,t) z^k P_n(x)
\end{equation}
with
\begin{equation}
q(z,t) = \frac{1}{z} + \sum_{l} q_l(t) z^l \\; \;
p(z,t) = \frac{1}{z} + \sum_{l} p_l(t) z^l
\end{equation}
Furthermore,
\begin{equation}
V_1'(x) P_{n-l}(x) = \sum_m C_{l m} P_{n-m}(x)
\end{equation}
where the coefficients $ C_{l m} $ can be computed by WKB formula
\begin{equation}
C_{l m} = \oint \frac{dz}{2\pi \imath} z^{l-m-1} V_1'(p(z,t))
\end{equation}
and likewise for $ Q $. The $z $ integration contour surrounds all the
singularities of the integrand, which for the polynomial potential would reduce
to the residue at $ z=0 $ or, equivalently, at $ z = \infty $.

Using these formulas and orthogonality conditions we find
\begin{equation}
q_l(t) = f^l(t) \oint \frac{dz}{2\pi \imath} z^{l-1} V_1'(p(z,t))
\label{Qfinal}
\end{equation}
\begin{equation}
p_l(t) = f^l(t) \oint \frac{dz}{2\pi \imath} z^{l-1} V_2'(q(z,t))
\label{Pfinal}
\end{equation}
and
\begin{equation}
f(t) = -t +  \oint \frac{dz}{2\pi \imath} \frac{V_1'(p(z,t))}{z^2}
\end{equation}

Multiplying (\ref{Pfinal}), (\ref{Qfinal}) by  $ y^l $ and summing over all $ l
=0,1, \dots \infty$ we arrive at the final form of the integral equations
\begin{equation}
q(y,t) = \frac{1}{y} + \oint \frac{dz}{2\pi \imath z}  \frac{V_1'(p(z,t))}{1- z
y f(t)}
\end{equation}
\begin{equation}
p(y,t) = \frac{1}{y} + \oint \frac{dz}{2\pi \imath z}  \frac{V_2'(q(z,t))}{1- z
y f(t)}
\end{equation}
where the integration contour encircles singularities of potentials,
 but excludes
the pole at $ z = \frac{1}{y f(t)} $. Computing residues in this pole
and at infinity, we arrive at algebraic form of equations
\begin{eqnarray}
f(t) &=& -t +Res \left[ \frac{V_1'(p(z,t))}{z^2} \right]_{z = \infty} \\
q(y,t) &=& \frac{1}{y} + V_1'\left(p\left(\frac{1}{yf(t)},t\right)\right) +
Res \left[ \frac{V_1'(p(z,t))}{z(1- z y f(t))} \right]_{z = \infty} \\
p(y,t) &=& \frac{1}{y} + V_2'\left(q\left(\frac{1}{yf(t)},t\right)\right) +
Res \left[ \frac{V_2'(q(z,t))}{z(1- z y f(t))} \right]_{z = \infty}
\end{eqnarray}

As for the free energy, in the WKB approximation we find
\begin{equation}
\frac{1}{N}\frac{\partial \ln I(\phi,\chi)}{\partial \phi_a} =  \int_0^1 dt
\oint \frac{dz}{2\pi \imath z} \frac{1}{p(z,t)-\phi_a}
\end{equation}

Let us  now compute the second derivatives of our integral with respect to
eigenvalues $ \phi_a , \chi_b $. The simplest one is the mixed derivative
\begin{equation}
\frac{\partial^2 \ln I(\phi,\chi)}{\partial \phi_a \partial \chi_b} =
\sum_{i j} \left\langle\frac{1}{(x_i-\phi_a)(y_j-\chi_b)} \right\rangle_N -
\sum_{i j} \left\langle\frac{1}{x_i-\phi_a} \right\rangle_N
\left\langle\frac{1}{y_j-\chi_b} \right\rangle_N
\end{equation}
where averaging $ <A>_N $ corresponds to initial measure, depending of all $ N
$ $ x, y $ variables, rather than above averaging $ <A> $ over just one pair.

In virtue of symmetry of the measure all terms with $ i \neq j $ contribute the
same as the term with $i=1,j=2$, and all terms with $ i=j $ contribute the same
as $ i=j=1$, therefore
\begin{equation}
\sum_{i j} \left\langle\frac{1}{(x_i-\phi_a)(y_j-\chi_b)} \right\rangle_N =
N(N-1) \left\langle\frac{1}{(x_1-\phi_a)(y_2-\chi_b)} \right\rangle_N + N
\left\langle\frac{1}{(x_1-\phi_a)(y_1-\chi_b)} \right\rangle_N
\end{equation}
\begin{equation}
\sum_{i j} \left\langle\frac{1}{x_i-\phi_a} \right\rangle_N
\left\langle\frac{1}{y_j-\chi_b} \right\rangle_N = N^2
\left\langle\frac{1}{x_1-\phi_a} \right\rangle_N
\left\langle\frac{1}{y_1-\chi_b} \right\rangle_N
\end{equation}

Repeating the same steps as before, when we derived (\ref{Hproduct}), we find
here after some cancellations
\begin{equation}
\frac{\partial^2 \ln I(\phi,\chi)}{\partial \phi_a \partial \chi_b} =
\sum_{n=1}^N \left\langle \frac{P_n(x_1)
\tilde{Q}_n(y_1)}{(x_1-\phi_a)(y_1-\chi_b)} \right\rangle -
\sum_{n,m=1}^N \left\langle \frac{P_n(x_1) \tilde{Q}_m(y_1)}{x_1-\phi_a}
\right\rangle
 \left\langle \frac{P_m(x_2) \tilde{Q}_n(y_2)}{y_2-\chi_b} \right\rangle
\end{equation}
where angular brackets already correspond to one dimensional averaging.

Now we further reduce terms by applying the completeness relation
\begin{equation}
\sum_{m=1}^{\infty}
\left\langle \frac{P_n(x_1) \tilde{Q}_m(y_1)}{x_1-\phi_a}\right\rangle
\left\langle \frac{P_m(x_2) \tilde{Q}_n(y_2)}{y_2-\chi_b} \right\rangle =
\left\langle \frac{P_n(x_1) \tilde{Q}_n(y_1)}{(x_1-\phi_a)(y_1-\chi_b)}
\right\rangle
\end{equation}
and we find after changing summation variables $ n = N-k,m=N+l+1 $
\begin{equation}
\frac{\partial^2 \ln I(\phi,\chi)}{\partial \phi_a \partial \chi_b} =
\sum_{k=0}^{N-1} \sum_{l=0}^{\infty}\left\langle \frac{P_{N-k}(x_1)
\tilde{Q}_{N+l+1}(y_1)}{x_1-\phi_a} \right\rangle
 \left\langle \frac{P_{N+l+1}(x_2) \tilde{Q}_{N-k}(y_2)}{y_2-\chi_b}
\right\rangle
\end{equation}
So far we did not use any approximations. Now we use the WKB approximation at
large $N$
\begin{equation}
\left\langle \frac{P_{N-k}(x_1) \tilde{Q}_{N+l+1}(y_1)}{x_1-\phi_a}
\right\rangle \rightarrow
\oint \frac{dz_1}{2\pi \imath} \frac{z_1^{k+l}}{p(z_1,1)-\phi_a}
\end{equation}
\begin{eqnarray}
\lefteqn{\left\langle \frac{P_{N+l+1}(x_2) \tilde{Q}_{N-k}(y_2)}{y_2-\chi_b}
\right\rangle =} \\ \nonumber
& & \frac{h_{N+l+1}}{h_{N-k}}\left\langle \frac{\tilde{P}_{N+l+1}(x_2)
Q_{N-k}(y_2)}{y_2-\chi_b} \right\rangle \rightarrow \\ \nonumber
& & f(1)^{k+l+1}\oint \frac{dz_2}{2\pi \imath}
\frac{z_2^{k+l}}{q(z_2,1)-\chi_b}
\end{eqnarray}

\begin{eqnarray}
\lefteqn{\frac{\partial^2 \ln I(\phi,\chi)}{\partial \phi_a \partial \chi_b} =}
\\ \nonumber
& &\oint \frac{dz_1}{2\pi \imath} \oint \frac{dz_2}{2\pi \imath}
\frac{f(1)}{(1-f(1)z_1z_2)^2 (p(z_1,1)-\phi_a)(q(z_2,1)-\chi_b)}
\end{eqnarray}

The computation of the double derivative $ \frac{\partial^2 \ln
I(\phi,\chi)}{\partial \phi_a \partial \phi_b} $ for $ a \neq b $ goes along
the same line untill the following point
\begin{eqnarray}
\lefteqn{\frac{\partial^2 \ln I(\phi,\chi)}{\partial \phi_a \partial \phi_b}
\stackrel{a \neq b}{=}} \\ \nonumber
& & \sum_{k=0}^{N-1} \sum_{l=0}^{\infty}\left\langle \frac{P_{N-k}(x_1)
\tilde{Q}_{N+l+1}(y_1)}{x_1-\phi_a} \right\rangle
\left\langle \frac{P_{N+l+1}(x_2) \tilde{Q}_{N-k}(y_2)}{x_2-\phi_b}
\right\rangle
\end{eqnarray}
When we apply the WKB approximation, we have to expand $ P $ polynomials in
both averages, unlike before, when we expanded $P$ in first average, and $Q$ in
the second one. The result is therefore, slightly different
\begin{equation}
\frac{\partial^2 \ln I(\phi,\chi)}{\partial \phi_a \partial \phi_b} \stackrel{a
\neq b}{=}
\oint \frac{dz_1}{2\pi \imath} \oint \frac{dz_2}{2\pi \imath}
\frac{1}{(z_1-z_2)^2 (p(z_1,1)-\phi_a)(p(z_2,1)-\phi_b)}
\end{equation}

At $ a=b$ there is an extra term, calculable by the same method,
adding this term we find
\begin{eqnarray}
\lefteqn{\frac{\partial^2 \ln I(\phi,\chi)}{\partial \phi_a \partial \phi_b} =}
\\ \nonumber
 & & N \delta_{a b} \int_0^1 dt \oint \frac{dz}{2 \pi \imath
z}\frac{1}{(q(z,t)-\phi_a)^2} \\ \nonumber
& & + \oint \frac{dz_1}{2\pi \imath} \oint \frac{dz_2}{2\pi \imath}
\frac{1}{(z_1-z_2)^2 (p(z_1,1)-\phi_a)(p(z_2,1)-\phi_b)}
\end{eqnarray}
However, as we found, the $ \delta_{ab} $ terms in the effective
Action all cancel among themselves. This $ \delta_{ab} $ term reduces
to $ V''(\phi) $ which adds to similar term from the Vandermonde
determinant and the mass term and cancel in virtue of the classical equation.

Let us discuss the choice of the $z $ integration contours. Originally they
enclosed the singularities of $ q(z,t) $ and all the zeros of denominators.
After summing the geometric series the new double poles appear. The pole at $
f(1) z_1 z_2 = 1 $ must be taken outside the integration contours, as it arose
after continuation of the geometric series from domain of convergence $ z_1 z_2
\rightarrow 0 $. The situation with the pole at $ z_1 = z_2 $ is slightly more
delicate. Never mind how small a circle you take for the $ z_1 , z_2 $
 integrations
this pole would always be located directly on this circle. However, the residue
at this pole, say, for $ z_2 $
\begin{equation}
\oint \frac{dz_1}{2\pi \imath}
\frac{1}{(p(z_1,1)-\phi_a)} \frac{\partial}{\partial z_1}
\frac{1}{(p(z_1,1)-\phi_b)}
\end{equation}
would be antisymmetric  with respect to $ a,b$, as the closed loop integral of
the total derivative vanishes. Hence, this pole should be disregarded, to
preserve the symmetry of the matrix of the second derivatives. In other words,
this is the principal value
\begin{equation}
\frac{1}{(z_1-z_2)^2} \rightarrow \frac{1}{2} \frac{1}{(z_1-z_2+ \imath 0)^2} +
\frac{1}{2} \frac{1}{(z_1-z_2- \imath 0)^2}
\end{equation}

\end{document}